\documentclass[british]{article}
\usepackage[T1]{fontenc}
\usepackage[latin9]{inputenc}
\usepackage{babel}
\usepackage{array}
\usepackage{textcomp}
\usepackage{amstext}
\usepackage{graphicx}
\usepackage[unicode=true]
 {hyperref}

\makeatletter

\providecommand{\tabularnewline}{\\}

\usepackage{url}

\usepackage{textcomp}
\newcommand{\microstrain}{\ensuremath{\,\mathrm{\mu{}e}}}
\newcommand{\mm}{\ensuremath{\,\mathrm{mm}}}
\newcommand{\g}{\ensuremath{\,\mathrm{g}}}
\newcommand{\dB}{\ensuremath{\,\mathrm{dB}}}
\newcommand{\dBi}{\ensuremath{\,\mathrm{dBi}}}
\newcommand{\amphour}{\ensuremath{\,\mathrm{Ah}}}
\newcommand{\volts}{\ensuremath{\,\mathrm{V}}}

\newcommand{\W}{\ensuremath{\,\mathrm{W}}}

\newcommand{\m}{\ensuremath{\,\mathrm{m}}}
\newcommand{\GHz}{\ensuremath{\,\mathrm{GHz}}}
\newcommand{\MHz}{\ensuremath{\,\mathrm{MHz}}}
\newcommand{\kHz}{\ensuremath{\,\mathrm{kHz}}}
\newcommand{\tc}{\ensuremath{\,\textrm{\textcelsius}}}

\makeatother

\begin{document}

\title{Proof of Concept of Wireless TERS Monitoring }

\author{Michael Allen\and{}Elena Gaura\and{}Ross Wilkins\and{}James Brusey\thanks{Corresponding
author: FMFM Research Centre, Coventry University, Priory Street,
Coventry CV1 5FB, UK E-mail: j.brusey@coventry.ac.uk}\and{}Yuepeng
Dong\thanks{Singapore-MIT Alliance for Research and Technology, Singapore
138602}\and{}Andrew J. Whittle\thanks{Department of Civil and Environmental
Engineering, MIT, Cambridge, MA 02142, USA}}

\date{11-July-2017 (author uncorrected pre-print)\thanks{Published version
is available from \href{http://dx.doi.org/10.1002/stc.2026}{DOI: 10.1002/stc.2026}}}
\maketitle
\begin{abstract}
Temporary earth retaining structures (TERS) help prevent collapse
during construction excavation. To ensure that these structures are
operating within design specifications, load forces on supports must
be monitored. Current monitoring approaches are expensive, sparse,
off-line, and thus difficult to integrate into predictive models.
This work aims to show that wirelessly connected battery powered sensors
are feasible, practical, and have similar accuracy to existing sensor
systems. We present the design and validation of ReStructure, an
end-to-end prototype wireless sensor network for collection, communication,
and aggregation of strain data. ReStructure was validated through
a six months deployment on a real-life excavation site with all but
one node producing valid and accurate strain measurements at higher
frequency than existing ones.  These results and the lessons learnt
provide the basis for future widespread wireless TERS monitoring that
increase measurement density and integrate closely with predictive
models to provide timely alerts of damage or potential failure. 
\end{abstract}

\section{Introduction}

Underground construction of subway stations and building basements
are frequently accomplished through `bottom-up' methods; where the
permanent structure is built within an excavation pit that is supported
by a Temporary Earth Retaining System (TERS). Typically TERS comprise
of perimeter walls that are supported internally by cross-lot bracing
struts, or externally by tieback anchors (installed in the adjacent
ground). These temporary structures are designed to support the loads
exerted by the retained soil mass (including pore water pressures),
forces induced by foundations from adjacent buildings, etc. Numerical
analyses of soil-structure interactions (usually by finite element
methods) are used to predict the performance of the structure including
the loads in the structural elements, magnitudes of wall deflections
and ground deformations at all stages of the construction process.
During construction, TERS systems are extensively monitored to ensure
the stability of the works and to control/mitigate impacts of excavation-induced
ground movements that can affect adjacent structures and facilities.
Field monitoring typically includes measurements of compression forces
in the strutting system (via strain gauges or load cells), deflections
(and flexure) of the retaining wall (via inclinometers), surface and
subsurface ground movements (optical surveys, inclinometers and extensometers)
and pore water pressures (piezometers). In current practice, the monitoring
data are primarily used to i) compare against the expected values
predicted by prior analyses of soil-structure interaction, to determine
if the design specification has been exceeded and ii) trigger static
threshold based flags/alarms. The high unit costs for installing these
instruments (particularly subsurface sensors) leads to sparse spatial
coverage at most construction sites. Furthermore, there is little
use of the data to interpret the actual field performance and hence,
to update model predictions and reduce risks associated with the construction
process. 

Wireless sensing offers an attractive alternate approach for monitoring
structural performance. By reducing the unit cost of sensor installation,
wireless sensor networks can increase the spatial and temporal resolution
for measuring forces in the bracing system (compared to wired devices).

The goal of this research is to enable robust, reliable wireless collection
of strain measurements in an excavation environment and make it available
for on-line integration, in real-time, with a user application. This
paper demonstrates:
\begin{itemize}
\item the feasibility of using Wireless Sensor Networks (WSNs) for TERS.
An end-to-end proof-of-concept system (ReStructure) has been developed
and deployed on an excavation site (alongside a conventional wired
system, where forces were monitored by vibrating wire strain gauges,
done by a specialist instrumentation contractor) to acquire, transmit
and aggregate strain data. Criteria used for evaluating the suitability
of the prototype for the application have been: i) the validity and
accuracy of the measurements against the industry gold-standard systems,
ii) wireless communication reliability and data yield, iii) system
survivability in field, and iv) system and battery life.
\item The paper provides insights into the feasibility of large scale, low
cost intelligent WSNs monitoring for TERS through analysis of the
data archive created through the pilot deployment. Lessons are drawn
from the system design, implementation and deployment that add to
the existing body of knowledge on deployable WSNs for structural monitoring.

\end{itemize}
In this paper we review the current literature (Section~\ref{sec:Literature-review})
on related wireless systems for structural health monitoring. We then
discuss the design specifications for the current application to Temporary
Earth Retaining systems (Section~\ref{sec:spec}), and present details
of the prototype design and implementation (Section~\ref{sec:WSN-System-design}).
We review data from two relatively short-duration field deployments
(Section~\ref{sec:Field-deployments}) and the lessons learned from
this experience (Section~\ref{sec:Lessons-from-the}).

\section{ReStructure: design specification and model integration\label{sec:spec}}

\begin{figure*}
\includegraphics{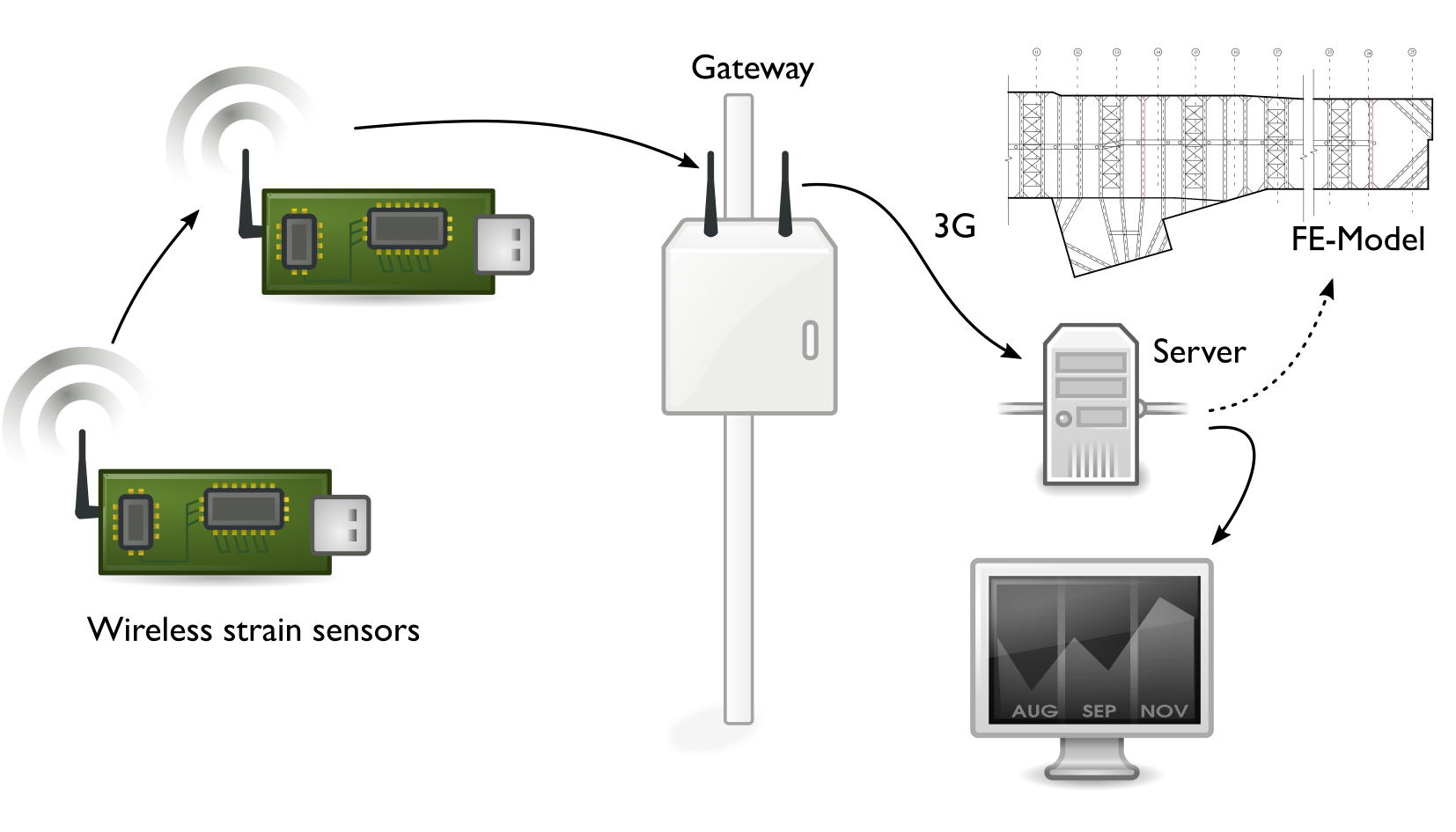}

\protect\caption{The ReStructure system architecture involves wireless strain sensors
transmitting their data locally to a gateway that then forwards via
3G to a server. The server can then be used to display or possibly
to integrate with a finite element (FE) model.\label{fig:Sensor-Network-Design}}
\end{figure*}

The goal of ReStructure is to enable collection of strain measurements
in an excavation environment and make it available for on-line integration,
in real-time, with a user application. Figure~\ref{fig:Sensor-Network-Design}
presents the system architecture.

\subsection{The WSN design specification}

The design specification for the WSN was developed to meet a variety
of constraints associated with the expected physical loads (strains),
project duration, sampling rate and unit costs of the sensor nodes.The
following generic requirements were defined at the outset:
\begin{description}
\item [{Data~acquisition~and~wireless~communication}] The sensor network
is expected to collect and transmit data at 5 minute intervals. The
nodes are powered via batteries.
\item [{Measurement~parameters}] The primary measurand is strain (as a
proxy for compressive load in the cross-lot bracing struts). Traditionally,
strain is measured using vibrating wire gauges or resistive foil gauges.
In commercial structural engineering projects, the standard is to
use vibrating wire gauges that are sampled at hourly intervals (or
less frequently) and have been shown to provide stable measurements
over long-term deployments in harsh environmental conditions. The
current deployment uses resistive foil gauges. These provide a small
form factor, have much lower unit costs and simpler field installation
(attachment by spot welding) but raise concerns about long term drift
in the measurements.. Environmental air temperature and humidity are
also measured in the proximity of the strain gauges in order to compensate
for environmental effects. The current application uses a single strain
gauge for axial compression of the strut (and assumes that bending
strains are negligible).
\item [{Target~lifetime}] The sensor network is intended to last for the
lifetime of an excavation (approximately 1 year). Individual nodes
are to be added to the network as the excavation progresses.
\item [{Packaging~and~deployment}] The strain gauges are mounted directly
on the struts. Sensor nodes must be attached close to the foil gauges
to reduce lead length and help protect the gauges. It is intended
for sensor nodes to be i) low cost to enable dense deployments and
ii) small in size and robust to prevent them being destroyed or stolen
during operation.
\end{description}
There are a number of challenges associated with devising a system
that fulfils this set of requirements:
\begin{enumerate}
\item maintaining an evolving and growing network in a climatically harsh
and dynamic construction environment 
\item meeting the system target life in face of environmental obstructions
(i.e, where transmission retries penalise the energy consumption)
\item ensuring sufficiently high data yield to enable its use in models
and user applications
\item ensuring adequate measurement accuracy with low cost sensing devices
\end{enumerate}
The lack of opportunity to retrieve and debug or otherwise access
nodes after installation is an added constraint and one that hindered
learning from node failure events.

\section{ReStructure: implementation and prototype characterisation\label{sec:WSN-System-design}}

To reduce risk and development time, the implementation was based
on off-the-shelf hardware and software wherever possible. A number
of custom hardware interfaces and software components were developed
in-house as specified below. The node and gateway energy profiling
and the network coverage / yield characterisation performed show the
alignment with the design requirements and open areas for further
work.

\subsection{Sensor node\label{sub:Sensor-node}}

\begin{figure}
\begin{centering}
\includegraphics[width=0.55\columnwidth]{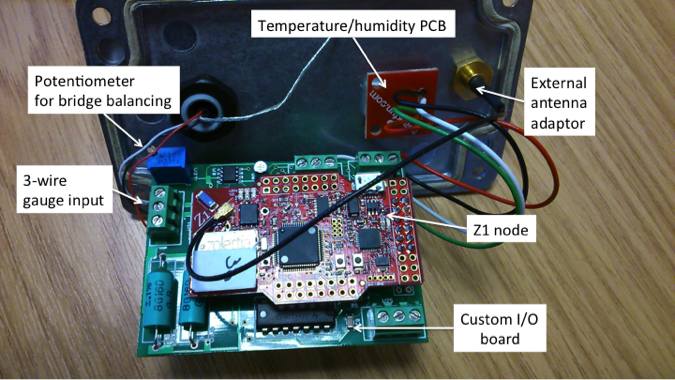}
\includegraphics[width=0.3\columnwidth]{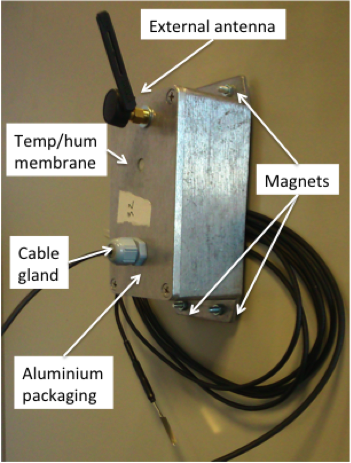}
\par\end{centering}

\protect\caption{ReStructure wireless sensor node hardware\label{fig:Restructure-Sensor-Node}}
\end{figure}

The sensor node (shown in Figure~\ref{fig:Restructure-Sensor-Node})
combines the Zolertia Z1 (MSP430 CPU and a CC2420 radio) platform
with a custom strain gauge PCB. The custom PCB supports one bonded
foil strain gauge whose readings are acquired using a Wheatstone bridge
combined with a low power 16-bit ADC (TI ADS1115). Measurement resolution
is $<1\microstrain$ with a measurement range of $\pm2\,500\microstrain$.
External temperature/humidity monitoring is enabled by a Sensirion
SHT15. Each sensor node is packaged in an IP65 aluminium enclosure
($115\times65.5\times50\mm$, $370\g$) with holes for an external
radio antenna ($4.4\dBi$ Antenova Titanis), a cable gland for the
strain gauge wiring, and a waterproof breathable membrane for the
SHT15. Magnetic feet on the enclosures allow the nodes to be placed
on the strut without welding. A deployed sensor node is shown in
Figure~\ref{fig:Example-deployment-on}.

The node's software was developed using Contiki WSN OS, which provides
a network stack with a low power MAC (ContikiMAC) and multi-hop tree
formation and data collection protocol (Contiki Collect). We reduced
the size of the recent message buffer kept by each node to minimise
the time taken to verify network formation after a node restart. Custom
drivers were written for the strain interface board.

At application level, a traditional sense and send system was implemented
that acquires, stores to flash, and transmits network information
(time, RSSI, beacon interval, sequence number, neighbours) as well
as a single strain and temperature/humidity measurement at 5 minute
intervals (92 bytes of payload total). Storing samples to flash means
that, even if the network fails, all measurements will be recorded
while the node has sufficient battery power. Oscilloscope--based energy
profiling was used to estimate the lifetime of the sensor nodes given
a particular battery capacity and sampling rate~\cite{brusey-energy}. 

\subsection{Gateway\label{sub:Gateway}}

\begin{figure}
\begin{centering}
\includegraphics[height=0.25\paperheight]{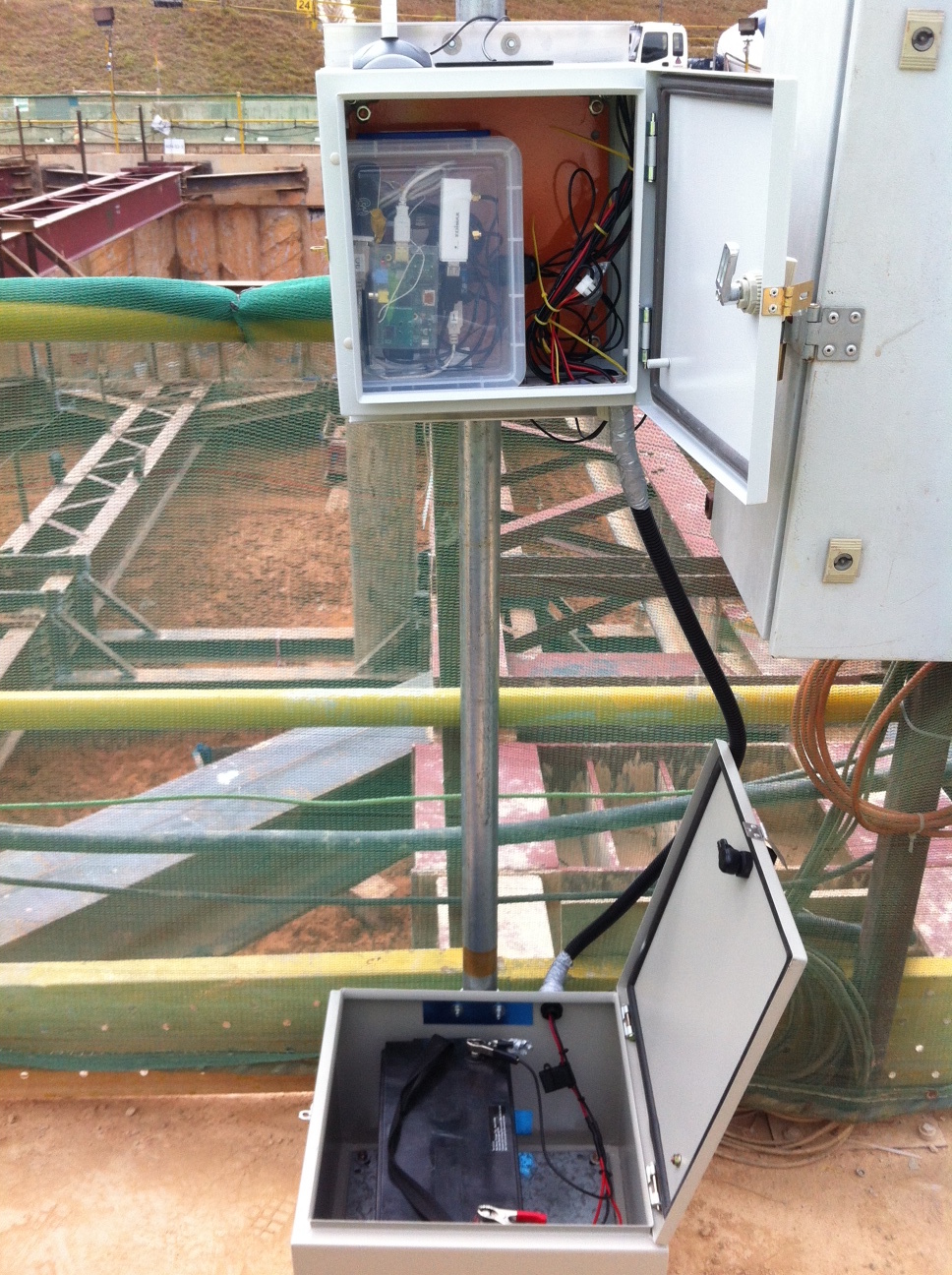}
\par\end{centering}

\protect\caption{Deployment of server at the construction site, nodes are placed at
the second strut level (approx. $12\m$ away)\label{fig:Server}}
\end{figure}

The Gateway was built using a Raspberry Pi model A+ combined with
a TelosB node and a USB 3G modem (both with external antennas). The
Gateway is deployed inside an IP65 mild steel enclosure and mounted
on a pole. Due to deployment constraints, the Gateway was powered
by a $12\volts$ $100\amphour$ battery (Figure~\ref{fig:Server}).
During normal operation WSN data is collected by the TelosB, aggregated
at the Raspberry Pi and transmitted hourly via 3G to a remote server
(and subsequently, compared with the FE model).

Based on prior experience with deploying WSNs, the Gateway design
and implementation focused on:
\begin{itemize}
\item minimising power consumption---by disabling HDMI on the Raspberry
Pi and devising custom circuitry to allow the power to the TelosB
and 3G modem to be controlled through GPIOs so that it can be turned
on only when needed. 
\item improving fault tolerance between the TelosB and the Raspberry Pi
to minimise on-site maintenance---by devising a simple handshake connection
so that the Raspberry Pi can periodically check and restart the TelosB
if necessary. For 3G transmissions, if an hourly update fails, it
is rolled into the following hour\textquoteright s transmissions,
and data is always archived locally in case of extended 3G outages.
\item ensuring on-site deployment, debugging and maintenance are as easy
as possible---the Gateway can host a USB WI-FI dongle. When the dongle
is connected, the Gateway becomes a wireless access point (using hostap~\cite{wikipedia12:_hostap}),
hosting a web page that shows real-time updates of data. Our measures
reduced the average power requirement of the gateway from over $4.8\W$
to $1.5\W$.
\end{itemize}
Based on gateway micro-benchmarking tests, we estimate that a $100\amphour$
battery will provide one month\textquoteright s operation (details
in Section~\ref{sub:Energy-consumption-benchmarking})

\subsection{Energy consumption benchmarking\label{sub:Energy-consumption-benchmarking}}

\paragraph{Node Energy}

\begin{figure}
\begin{centering}
\includegraphics{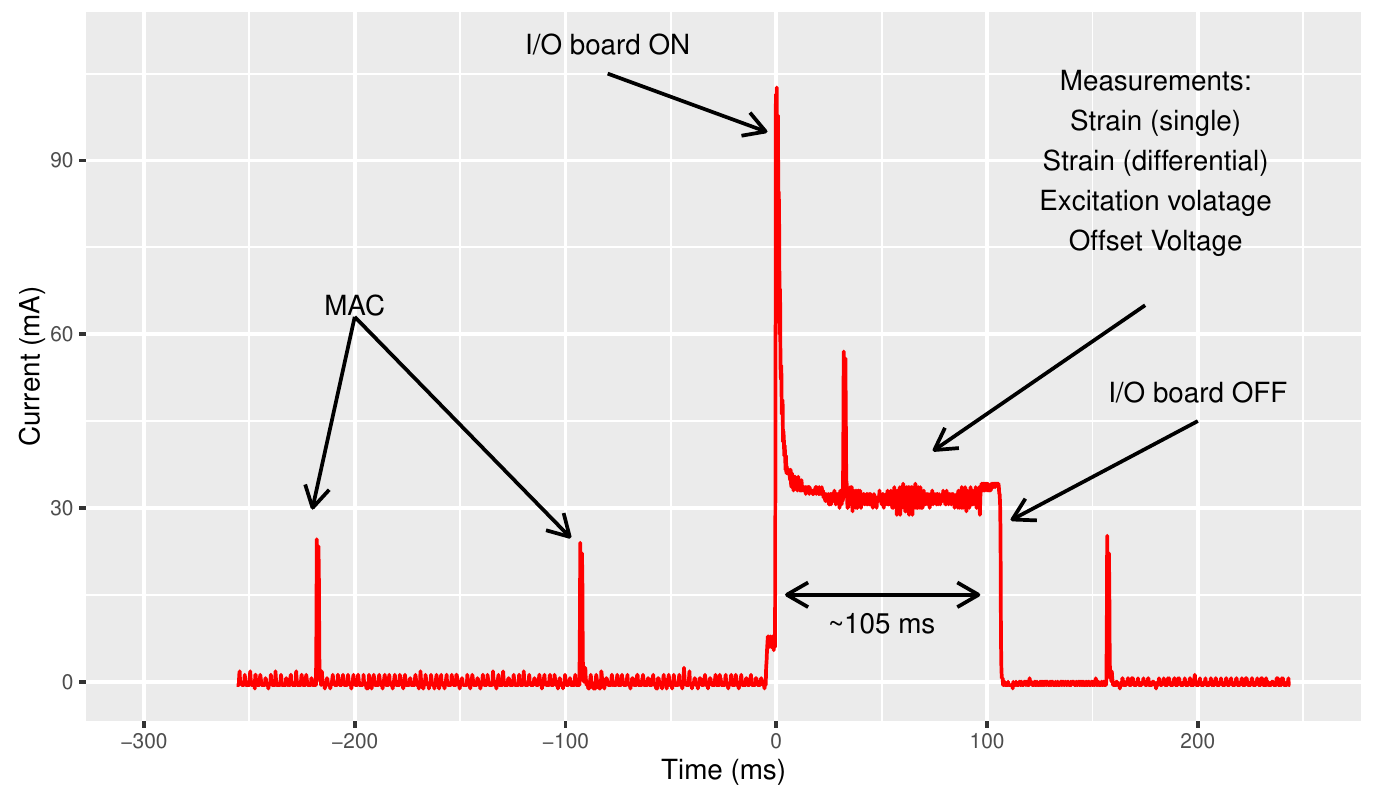}
\par\end{centering}

\protect\caption{The characteristic current consumption when switching on strain bridge
and taking a sample. Note that this profile does not include time
spent to \textquoteleft warm-up\textquoteright{} the bridge, i.e.
ensure that the strain value has reached a stable reading before taking
a measurement. \label{fig:StrainSample}}
\end{figure}

Benchmark measurements were conducted for the node operations and
used to predict lifetime based on $2\times\mathrm{AA}$ batteries.
Power was estimated by measuring the voltage drop over a shunt resistor
on the ground line of the power supply. The power consumption for
the following operations was determined:
\begin{itemize}
\item sense/sample strain only (no warm up time on bridge) 
\item sense/sample strain and temp/humidity 
\item sense/sample strain and write data to flash storage 
\item sense/sample strain and send data over network 
\item idle (no sampling, listening for packets)
\end{itemize}
Figure~\ref{fig:StrainSample} shows a typical power use profile
when switching on strain bridge and taking a sample. Each point on
the graph is an average of 64 oscilloscope samples. Based on these
measurements, we derived Table~\ref{tab:SS-node}, which shows an
estimated breakdown of power consumption by operation. 
\begin{table}
\begin{centering}
\protect\caption{Micro-benchmark estimates for a ReStructure node running sense-and-send.
The average current, for all but Idle, excludes the idle current.
\label{tab:SS-node}}

\par\end{centering}

\centering{}%
\begin{tabular}{cccc}
Process & Duration (ms) & Average current (mA) & mW\tabularnewline
\hline 
Warm up bridge & $5\,000$ & $33.6-1$ & $1.63$\tabularnewline
Strain sample & $105$ & $33.6-1$ & $0.034$\tabularnewline
Temp/humidity sample & $315$ & $33.6-1$ & $0.1$\tabularnewline
Write data to flash & $8$ & $6.7-1$ & $0.000\,46$\tabularnewline
Send message & $30$ & $17.9-1$ & $0.005\,1$\tabularnewline
Idle & $300\,000$ & $1$ & $3$\tabularnewline
\hline 
Total &  &  & $4.8$\tabularnewline
\end{tabular}
\end{table}

The energy benchmarking results in Table~\ref{tab:SS-node} provide
a profile of the design and help to identify opportunities for improvement.
Note that the time taken to sample temperature and humidity is almost
three times the time taken to acquire the strain measurements. Because
the I/O board controls the temperature/humidity sensor\textquoteright s
power, the Wheatstone bridge is also powered during this time. A refinement
of the PCB design may add an extra switch to allow the Wheatstone
bridge to be powered independently of the temperature/humidity sensor. 

The individual operations of a sensing cycle (5 minute interval) were
measured and aggregated to determine a simple, best--case baseline
for node lifetime of 205~days of operation with $2\times7.8\amphour\times1.5\volts$
C-cells.

\paragraph{Gateway}

Table~\textbf{\ref{tab:gateway-bench}} shows an estimated breakdown
of power consumption by functionality. Using a $12\volts$ $100\amphour$
car battery we estimate the gateway life to be 31 days. Future designs
should consider making use of solar panels.

\begin{table}
\begin{centering}
\protect\caption{Micro-benchmark estimates for the ReStructure gateway.\label{tab:gateway-bench}}

\par\end{centering}

\centering{}%
\begin{tabular}{cccc}
Process & Duration (s) & Average energy (A) & As\tabularnewline
\hline 
3G data transmission & 60 & 0.27 & 16.2\tabularnewline
Plotting & 60 & 0.13 & 7.8\tabularnewline
Processing & 3480 & 0.126 & 438.48\tabularnewline
\hline 
Total & 3600 &  & 462.48\tabularnewline
\end{tabular}
\end{table}

\subsection{Network performance\label{sub:Network-operation}}

\begin{figure}
\begin{centering}
\includegraphics[width=1\columnwidth]{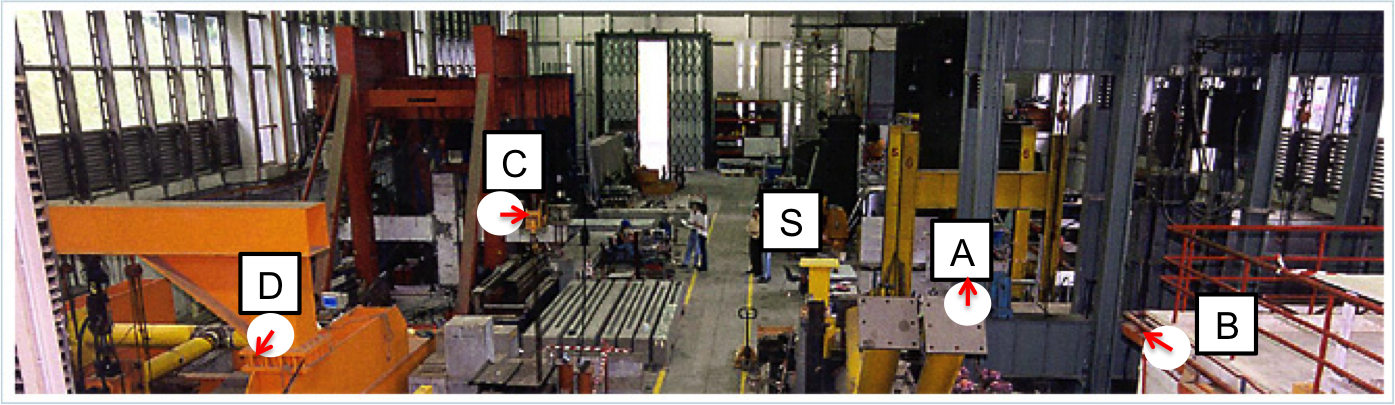}
\par\end{centering}

\protect\caption{Approximate layout of nodes during network test. Arrows indicate antenna
orientation; the sink is located roughly 3m to the left of node A.\label{fig:Approximate-layout}}
\end{figure}

The multi-hop network characterisation below provides insights into
the performance and yield expected from the ReStructure prototype
in a heavy machinery environment (see Figure~\ref{fig:Approximate-layout})
but without the harsh climatic conditions of the construction site.
Over a series of tests, each of the nodes transmitted \textquoteleft dummy\textquoteright{}
measurement packets at 5 minutes interval.

Table~\ref{tab:1-day-test} provides a summary of the performance
of the network over a (roughly) 23 hour test (transmitting about 276
packets). The performance is summarised by two measures, Packet Delivery
Ratio (PDR) and Link Stability. PDR is the percentage of packets received
vs. expected. We define \emph{link stability} as the proportion of
time that nodes are sending beacons at the longest interval ($3600\,\mathrm{s}$).
Collection tree protocols, such as Contiki Collect, broadcast fewer
beacon messages when the network is stable~\cite{Gnawali:2009:CTP:1644038.1644040}.
In addition, the number of packets sent directly to the sink node
versus other nodes is recorded to identify the proportional use of
direct versus multi-hop transmission.

\begin{table}
\centering{}\protect\caption{Summary of network statistics from 1-day multi-hop network deployment.\label{tab:1-day-test}}
\begin{tabular}{cc|cccc}
Node & Duration (h) & PDR (\%) & Link stability  & Most common  & MCP (\%)\tabularnewline
 &  &  & (\%) & parent & \tabularnewline
\hline 
A & 22.7 & 91.5 & 1.6 & Sink & 57.8\tabularnewline
B & 22.8 & 100.0 & 94.5 & Sink & 100.0\tabularnewline
C & 22.7 & 89.3 & 2.1 & Sink & 70.0\tabularnewline
D & 22.6 & 100.0 & 94.6 & Sink & 100.0\tabularnewline
\end{tabular}
\end{table}

Over the testing period, we observed two stable network connections,
with 100\% PDR, that were directly connected to the sink (Nodes B
and D), and two unstable network connections (Nodes A and C) that
had a lower PDR, were sending through two or more hops, and were changing
path frequently. This test indicates that that data yield and network
stability are related and that if stability is low, data yield will
also be low.

\section{Field deployment\label{sec:Field-deployments}}

\begin{figure}
\begin{centering}
\includegraphics[height=0.25\paperheight]{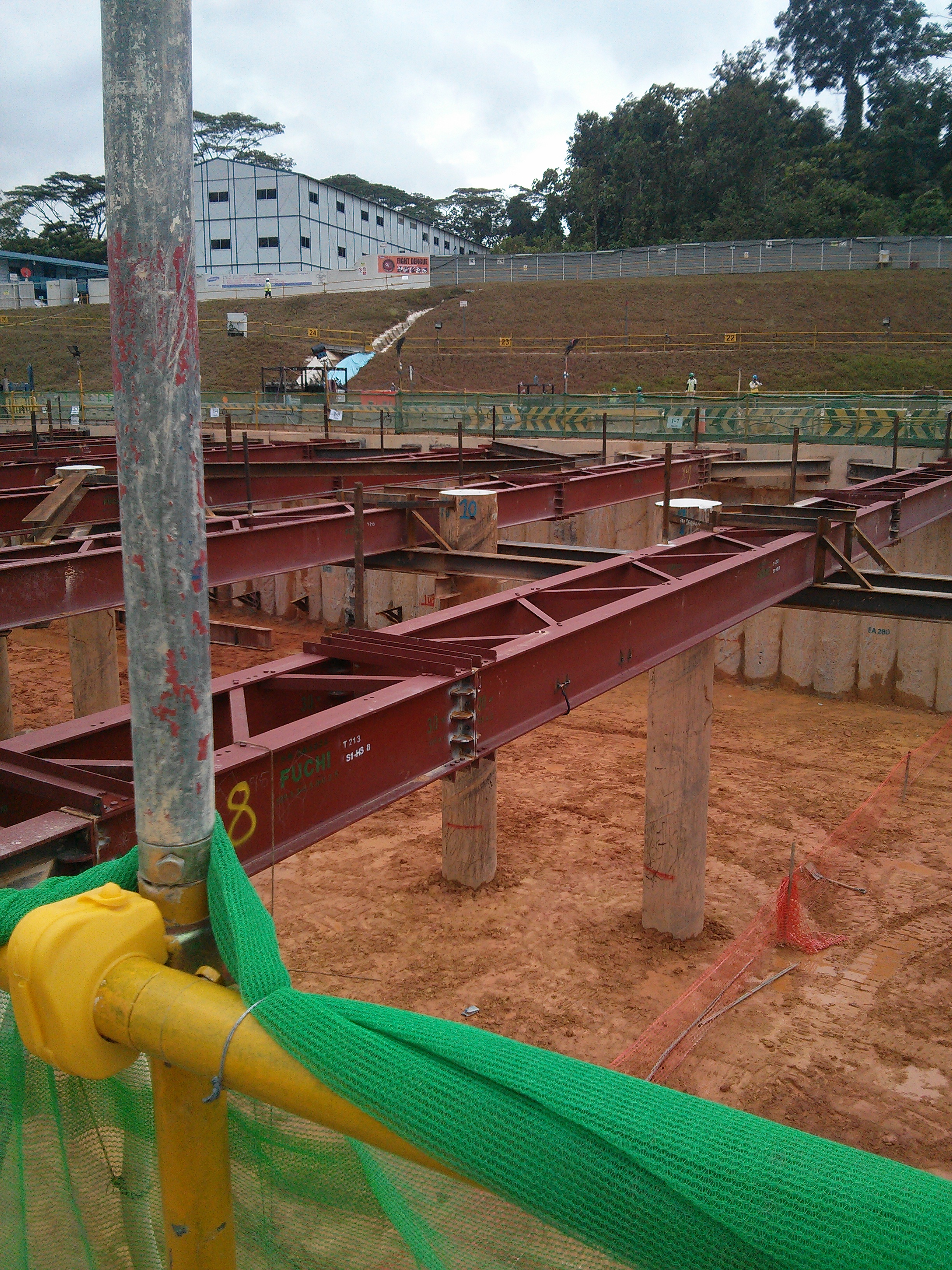}
\par\end{centering}

\protect\caption{Excavation pit being prepared for the insertion of Level 2 struts.\label{fig:Struts}}
\end{figure}

\begin{figure*}
\begin{centering}
\includegraphics[width=1\columnwidth]{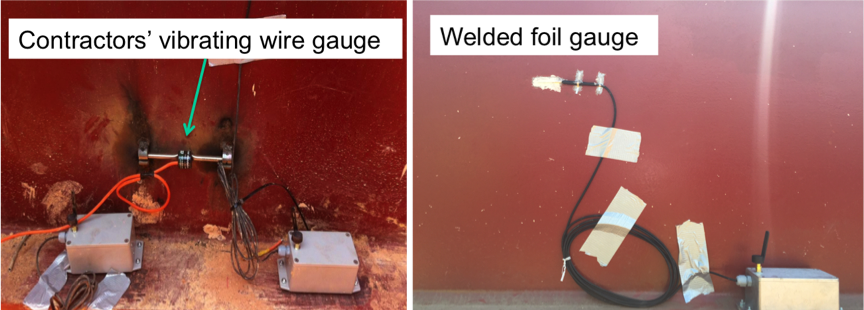}
\par\end{centering}

\protect\caption{Example deployment on a strut next to vibrating wire gauges\label{fig:Example-deployment-on}}
\end{figure*}

The prototype ReStructure system was deployed to monitor axial strains
for a series of selected struts within a large excavation pit. Data
was collected between May and December 2015. The deployment environment
was particularly challenging, due to: 
\begin{itemize}
\item \textbf{Climate} - The site is located in a tropical climate with
frequent and severe rainstorms. Temperatures can rise to $35\tc$
and 99\% RH above ground, and above $40\tc$ inside the excavation.
There are large daily cycles where temperature can change by $24\,\textrm{\textcelsius}$
and relative humidity by 60\%. Temperature and thus thermal expansion
affects both the TERS and the strain gauges. In addition, strain gauges
are subject to apparent strain, a change in resistance that is not
due to a loading change in the strut, but is related to the differences
in thermal expansion coefficient between the strain gauge and strut
being monitored.
\item \textbf{Obstructions} - Radio channel quality was expected to be low
due to the construction environment, which is mainly a combination
of metal and concrete materials that can reflect and attenuate radio
signals. The construction site itself is very dusty and there is lots
of machinery and movement that could damage equipment. This was expected
to affect the packet delivery radio (data yield), and therefore the
energy consumption due to retransmissions of messages. 
\item \textbf{Construction process} - A construction site is constantly
changing: over days and months material is being excavated from the
bottom of the trench and new struts and supports are being placed.
Therefore, the communication environment can be heavily affected as
struts are deployed, providing new surfaces that can reflect and interfere
with signals. 
\end{itemize}
Deployment of five nodes at three levels on a single strut line in
the excavation took place between 4th May and 18th July 2015. Level
1 deployment took place on 4th May 2015 (2 nodes), level 2 deployment
took place on 9th June 2015 (2 nodes) and level 3 deployment took
place on 18th Jul 2015 (1 node). Figures \ref{fig:Struts} and \ref{fig:Example-deployment-on}
show the deployment environment and an example of a node with a wireless
bonded foil (WBF) strain gauge strain gauge deployed next to the measurement
contractor's vibrating wire (VW) gauge.

The ReStructure nodes were mounted on to struts after it had been
placed in the excavation, but before it had been pre--loaded: installation
of the strut is completed by jacking preload forces into the strut
to ensure that it remains in compression during subsequent excavation
events. This allowed us to calibrate the WBF strain gauge in an unstressed/zero
strain configuration. The WBF strain gauges were spot welded aligned
to the length of the strut on the web (vertical section of strut,
see Figure~\ref{fig:Example-deployment-on}) and the strain bridge
was manually balanced by adjusting the on-board potentiometer while
the node is connected to a laptop displaying the current reading.
This balancing procedure allows a reference point to be set that corresponds
to zero load on the strut.

\begin{figure}
\begin{centering}
\includegraphics{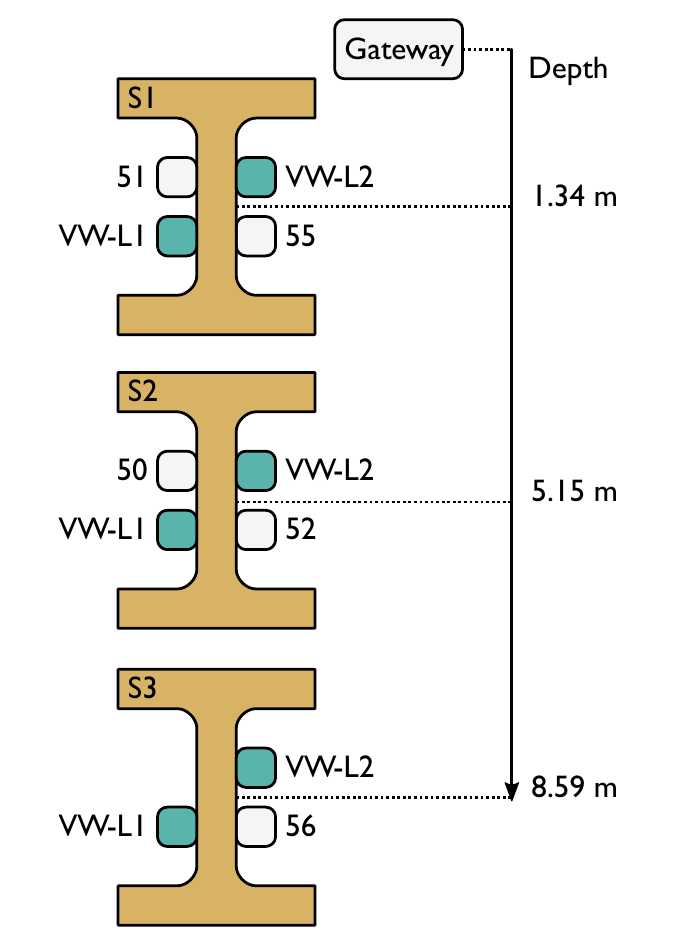}
\par\end{centering}

\protect\caption{\label{fig:strut-layout}Deployment layout of ReStructure wireless
bonded foil (WBF) strain gauges (50--56) and commercial vibrating
wire (VW) gauges (VW-L1/L2) across three strut levels (S1, S2, S3).
Distance from deployment location to gateway is shown in metres.}
\end{figure}

Figure~\ref{fig:strut-layout} shows a schematic elevation of the
cross section of the struts and the relative distance between the
nodes and the gateway. Data was collected at 5 minute intervals, and
nodes communicated on 802.15.4 channel 26 in an attempt to avoid any
WI-FI interference and the maximum number of retries for each packet
(sent by Contiki Collect) was set at 16. 

\subsection{Network and energy performance}

Table~\ref{tab:data-yield} shows the deployment duration and data
yield across the deployment. There were two major outages where the
node gateway batteries had exhausted: the first outage lasted 2.5
days (22nd to 24th June 2015) and the second outage lasted 19.3 days
(10th to 29th September 2015), due to difficulties accessing the site.
The data yield (or PDR) when ignoring the gateway outages is shown
in the far right colum in Table~\ref{tab:data-yield}.

Based on the analysis in Section~\ref{sub:Energy-consumption-benchmarking},
it was predicted that each node would have a best--case lifetime on
205 days. We see that both nodes on S2 got to within 10\% of this
prediction, where as the batteries in nodes on S1 and S3 lasted <50\%
less than expected. There are two potential explanations for this
discrepancy - firstly, the effect of radio retransmissions and packet
forwarding on lifetime (not analysed in this work), and secondly choice
of battery. Nodes 50 and 52 were fitted with Energizer MAX Alkaline
C cells, while the other nodes were fitted with Duracell PROCELL Alkaline
C cells, suggesting that the Energizer batteries performed better
for the energy consumption profile outlined in Section~\ref{sub:Energy-consumption-benchmarking}.

\begin{table}
\begin{centering}
\protect\caption{\label{tab:data-yield}Deployment summary for H14}

\par\end{centering}

\centering{}%
\begin{tabular}{cc|ccccc}
Strut  & Node  & Deployed & Offline & Duration  & Data  & Data yield\tabularnewline
level & ID &  &  & (days) & yield (\%) & w/o \tabularnewline
 &  &  &  &  &  & outages (\%)\tabularnewline
\hline 
S1 & 55 & 4th May & 16th Aug & 104 & 86.1\% & 88.2\%\tabularnewline
S1 & 51 & 4th May & 26th Jul & 82 & 91.1\% & 93.9\%\tabularnewline
S2 & 52 & 9th Jun & 15th Dec & 189 & 87.6\% & 99.0\%\tabularnewline
S2 & 50 & 9th Jun & 11th Dec & 185 & 87.4\% & 99.1\%\tabularnewline
S3 & 56 & 18th Jul & 4th Nov & 109 & 81.3\% & 98.7\%\tabularnewline
\end{tabular}
\end{table}

The Network configuration for the period 9th June to 4th November
2015, where all nodes were deployed and operating concurrently, is
shown in Table~\ref{tab:network-config}. The network performance
is evaluated using link stability, as defined in Section~\ref{sub:Network-operation},
as well as most common parent (MCP). The network was stable throughout
($>90$\% stability for each node), and most often at S1 and S2, nodes
chose to send directly to the sink. At S3, node 56 chose to send via
S2 most often.

\begin{table}
\centering{}\protect\caption{\label{tab:network-config}Network configuration from 9th June 2015
to 4th November 2015 (148 days)}
\begin{tabular}{cc|cccc}
Strut  & Node  & Dist. to  & Most common  & MCP \% & Stability\tabularnewline
level & ID & sink (m) & parent &  & \tabularnewline
\hline 
S1 & 55 & 1.34 & Sink & 98.4\% & 97.2\%\tabularnewline
S1 & 51 & 1.34 & Sink & 73.3\% & 92.7\%\tabularnewline
S2 & 52 & 5.15 & Sink & 86.7\% & 95.6\%\tabularnewline
S2 & 50 & 5.15 & Sink & 81.0\% & 93.9\%\tabularnewline
S3 & 56 & 8.59 & 50 & 64.6\% & 90.0\%\tabularnewline
\end{tabular}
\end{table}

\begin{figure}
\begin{centering}
\includegraphics{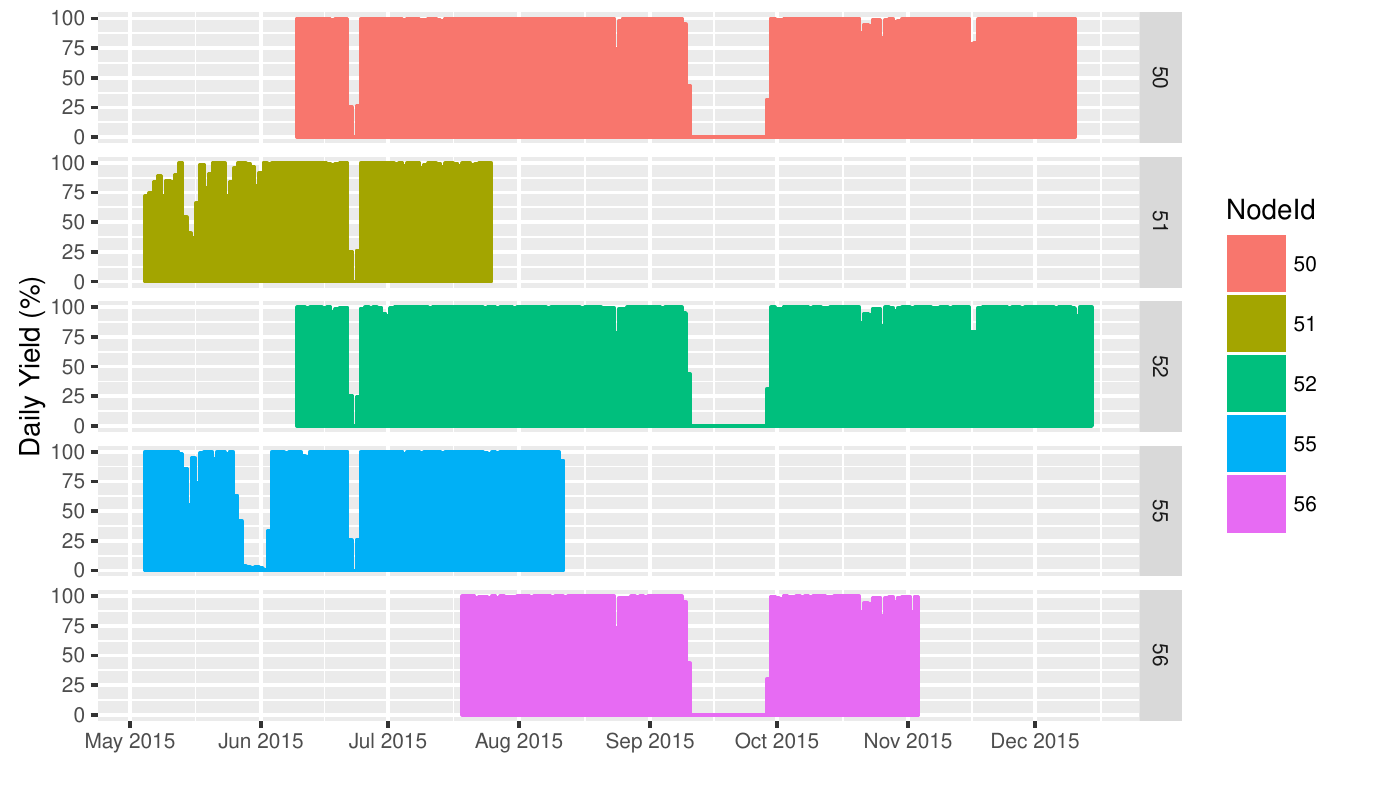}
\par\end{centering}

\protect\caption{Daily PDR for Deployment 2 showing some independent yield loss (e.g.,
late May for node 55) and some concurrent loss (e.g., September for
nodes 50, 52, 56). Concurrent losses are most likely due to gateway
failure.\label{fig:Heat-map}}
\end{figure}

Figure~\ref{fig:Heat-map} shows an annotated graph of the daily
PDR. In some cases, packet loss is specific to the node and is probably
due to localised wireless interference or shadowing. Most loss, however,
is concurrent indicating that the cause is most likely at the gateway.

\subsection{Strain measurement results}

\begin{figure}
\includegraphics{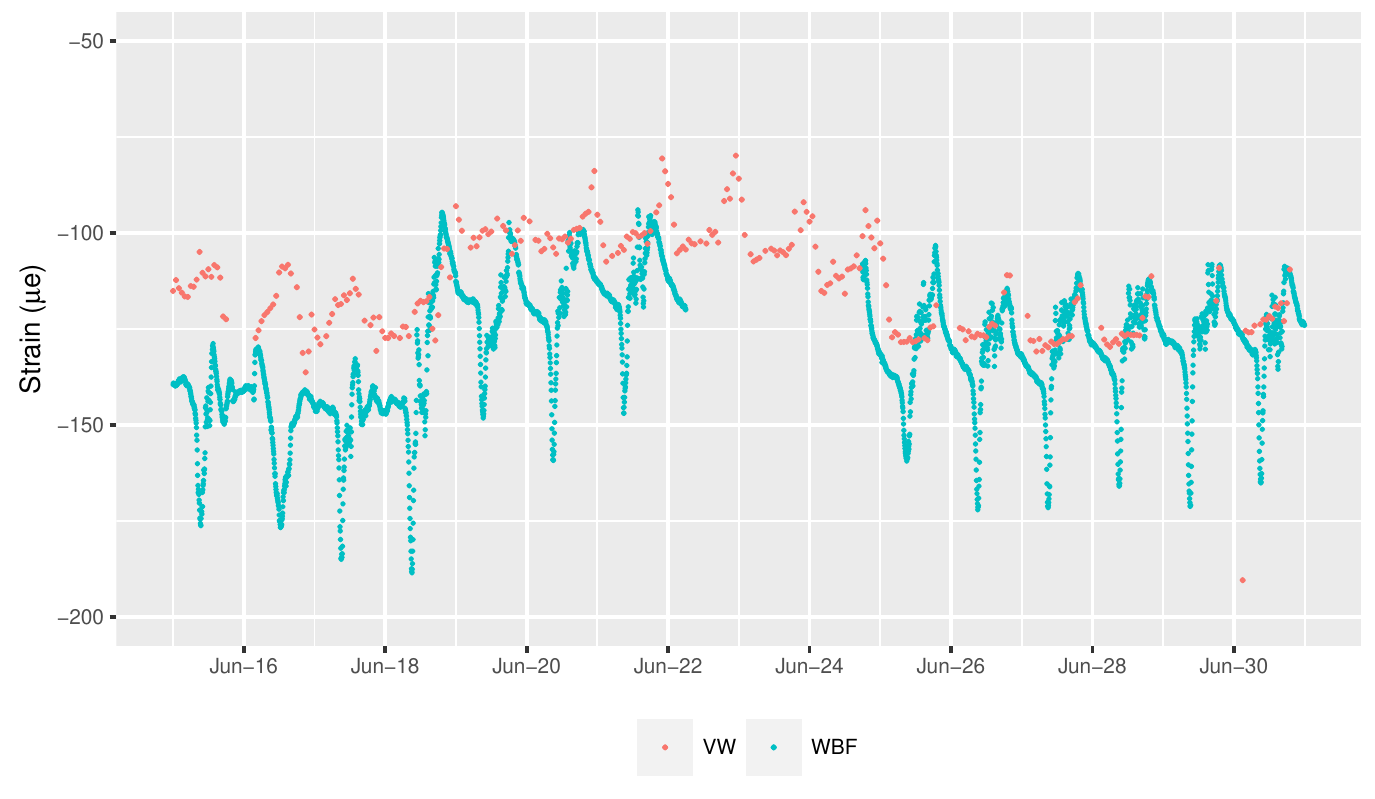}

\protect\caption{\label{fig:S1-data}Comparison of vibrating wire (VW) and wireless
bonded foil (WBF) strain measurements at S1-L1}
\end{figure}

WBF strain data was collected by the ReStructure nodes at 5 minute
intervals and VW strain data was collected by contractors at roughly
hourly intervals. 
\begin{itemize}
\item Figure~\ref{fig:S1-data} compares the VW strain measurements with
the WBF strain gauge. 
\item The VW measurements are compensated for temperature and hence, aim
to measure only the loads from the earth pressures. 
\item The VW data contain significant noise, skewed in the negative direction,
possibly associated with electromagnetic interference and which is
reduced in the figure by using a median filter ($k=5$). 
\end{itemize}

\begin{figure}
\includegraphics{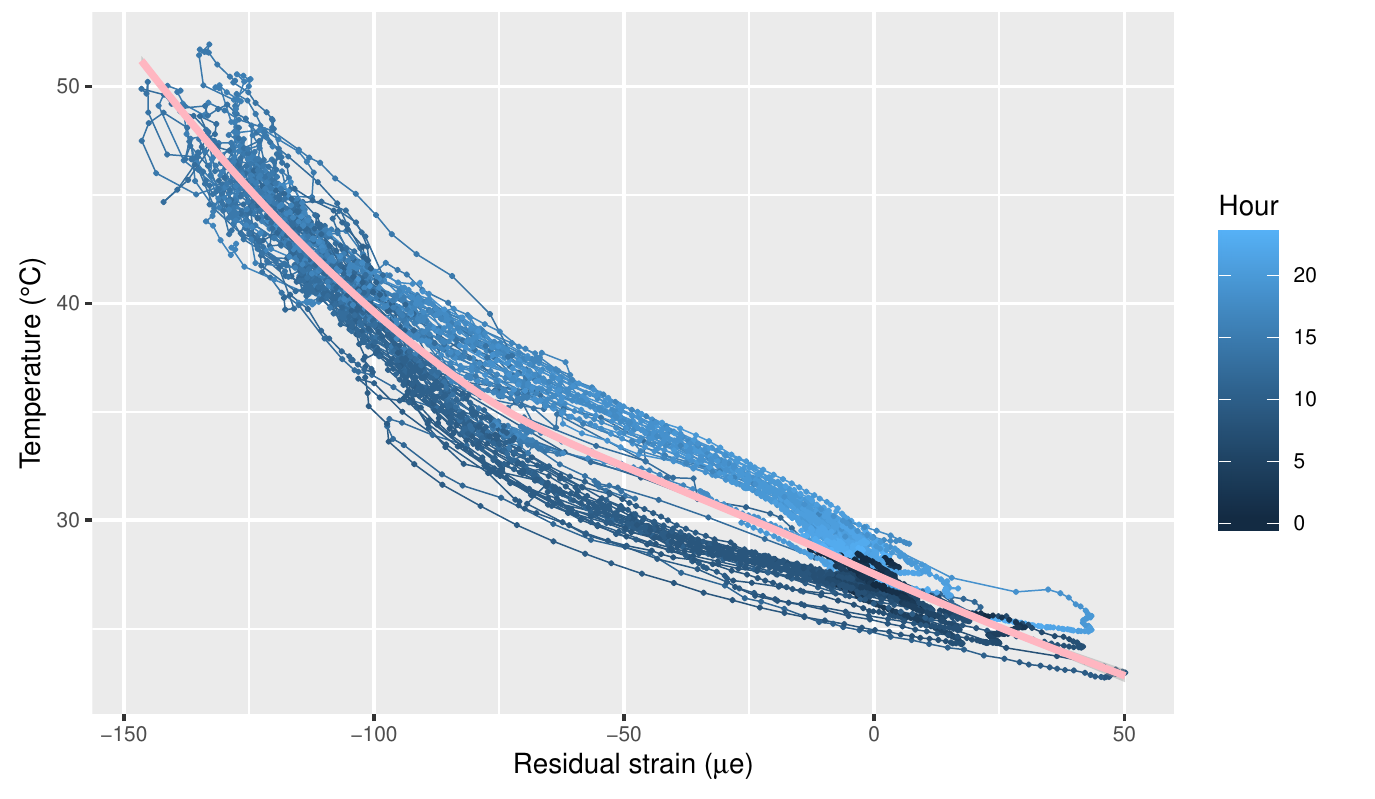}

\protect\caption{\label{fig:s1-temp-compensation}Residual strain versus temperature
providing thermal compensation curve for S1-L1 node 51 along with
a LOESS fit that was used to derive the compensated measurement shown
in Figure~\ref{fig:S1-data}. }
\end{figure}

\begin{itemize}
\item The WBF measurements could not be compensated using a factory polynomial
as none was provided by the manufacturer. Therefore, we used part
of the data to compensate under the assumption that temperature is
the main cause of diurnal variation.
\item To derive a thermal compensation curve for the WBF measurements, we
used a LOESS fit to establish the overall trend for strain where the
temperature was $27\pm0.1\,\tc$ and then found the residual strain
(for all temperatures) against this fit.

\begin{itemize}
\item This residual is then mapped to temperature as shown in Figure~\ref{fig:s1-temp-compensation}
to derive a temperature compensation curve.
\item Note that the effect on residual strain is larger (more negative)
for the same temperature when the temperature is increasing (during
the morning) than when the temperature is decreasing (during the evening). 
\item Compensating for temperature does not completely remove the diurnal
variation and there may be other factors (e.g., temperature at other
locations, rate of change of temperature, or the time of day) that
explain this variation. 
\end{itemize}
\end{itemize}

\begin{figure}
\includegraphics{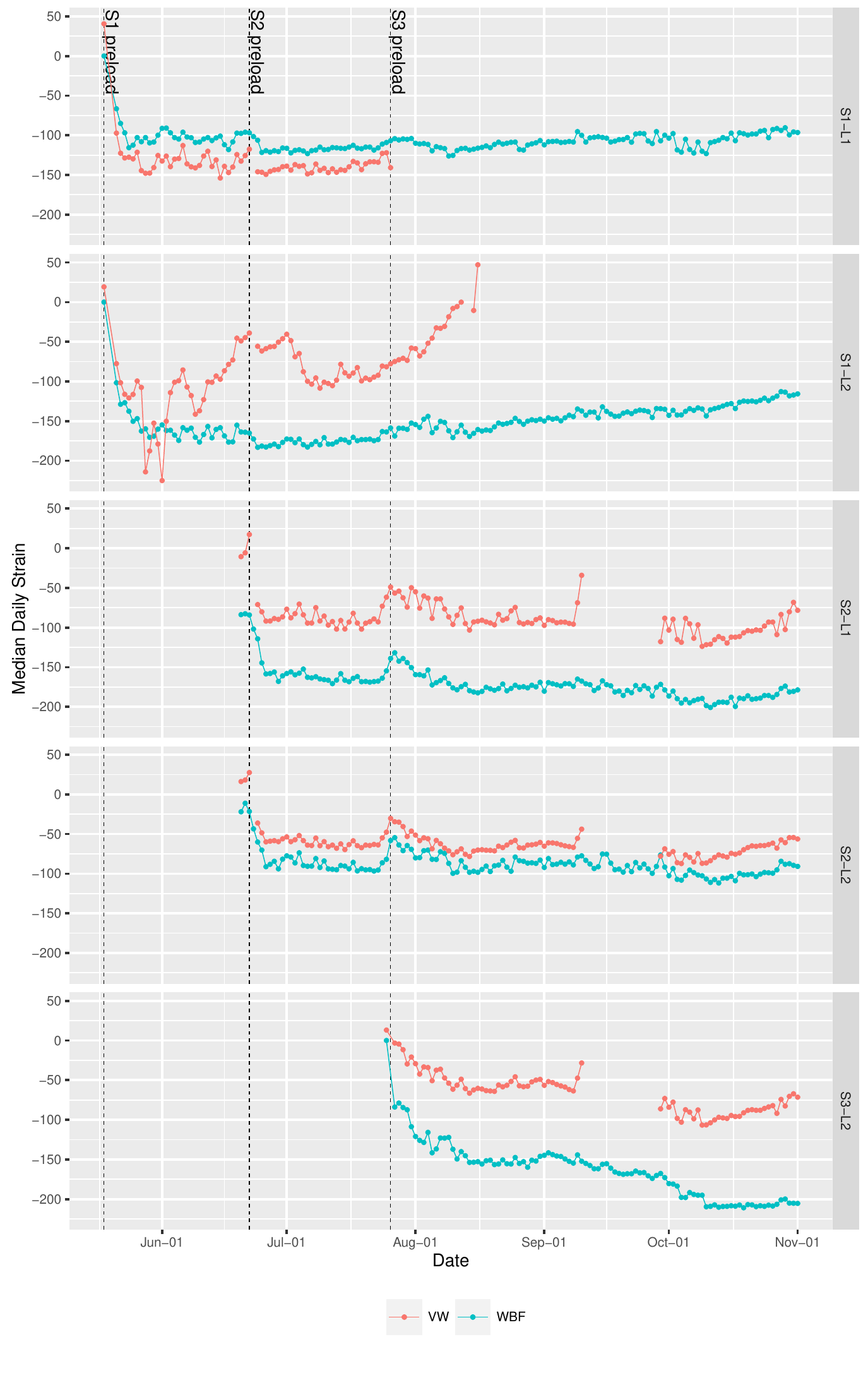}

\protect\caption{\label{fig:S1-median}Daily median strain comparing WBF with VW }
\end{figure}

\begin{itemize}
\item To allow comparison of the VW and WBF in terms of long-term trend,
we examined the daily median strain over time, as shown in Figure~\ref{fig:S1-median}.
\item As shown in Table~\ref{tab:median-correlation}, most WBF sensors,
with the exception of S1-L2 (Node 55), correlated well with the VW
sensor at the same site. 
\item Although it is not clear what the cause of failure was, node 55 may
have been affected by poor bonding of the foil to the strut. 
\item In the other WBF sensors, there is no drift in readings apparent. 
\item Correlated WBF sensors are offset from the corresponding VW sensors
and this offset is probably due to differences in calibration procedures
between the two. WBF sensors were zeroed prior to preloading whereas
VW sensors were zeroed during preloading.
\item Between stages, it is clear that the preloading phases for S2 and
S3 had an effect on the strain measurement for the struts above them. 
\end{itemize}

\begin{table}
\protect\caption{\label{tab:median-correlation}Pearson correlation and mean difference
(offset) for daily median strain of VW versus WBF over different sites}

\centering{}\begin{tabular}{cccc}
Site & Node & Correlation & Offset $\microstrain$\\
\hline
S1-L1 & 51 & 0.89 & -26 $\pm$ 3\\
S1-L2 & 55 & 0.19 & 81 $\pm$ 10\\
S2-L1 & 50 & 0.83 & 83 $\pm$ 2\\
S2-L2 & 52 & 0.91 & 26 $\pm$ 1\\
S3-L2 & 56 & 0.93 & 98 $\pm$ 4\\
\hline
\end{tabular}
\end{table}

\begin{figure}
\includegraphics[width=1\columnwidth]{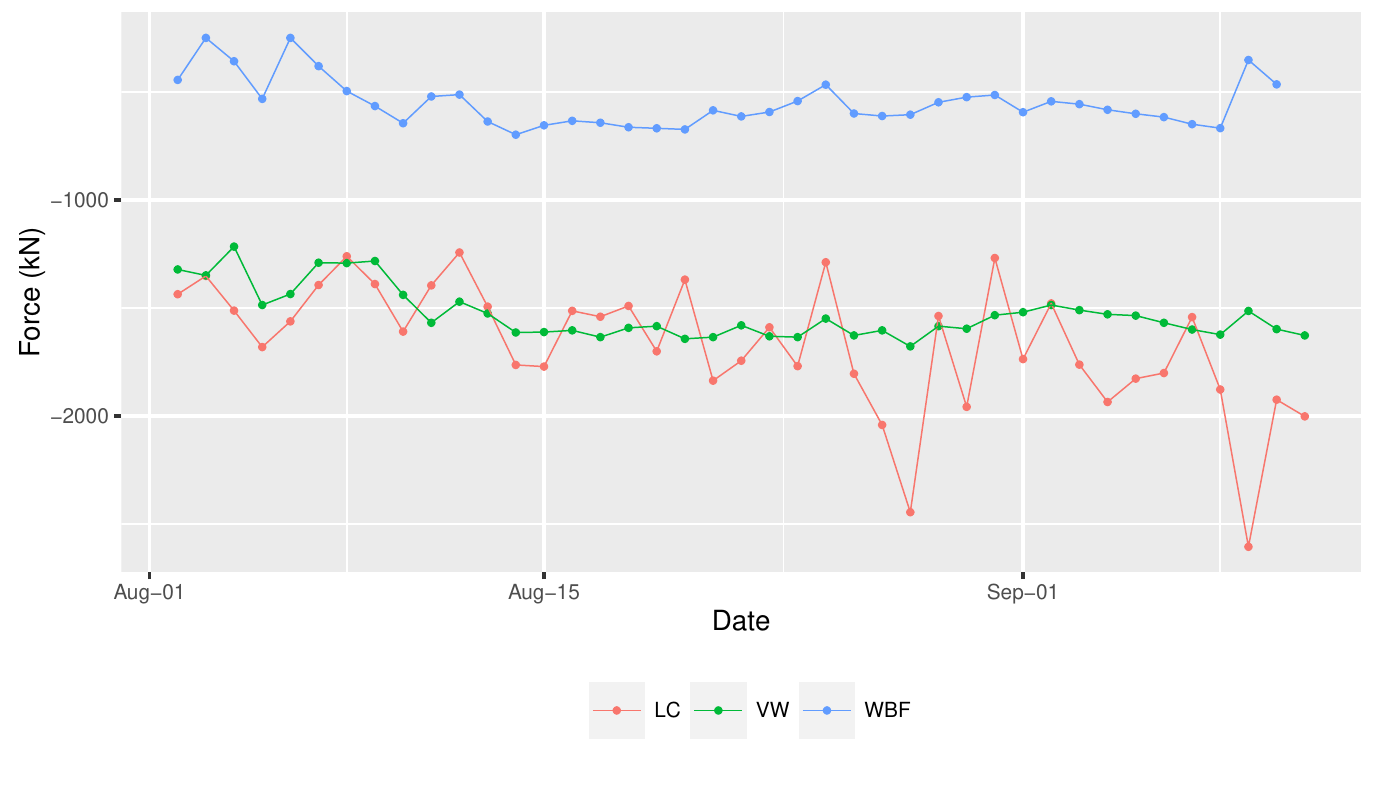}

\protect\caption{\label{fig:Load-data}ReStructure strain gauge matches load cell;
however both appear to be affected by temperature.}
\end{figure}

The contractor VW strain gauges are 'zeroed' at the time when the
pre-load is applied to the strut. Our WBF gauges were 'zeroed' at
some time prior to the preload, and we were not able to join during
the process. This means there is an offset between our WBF measurements
and the contractor VW measurements that is a function of the time
of day they were zeroed, in terms of apparent strain and thermal loading.
This offset problem is observable in all data, but particularly in
Figure~\ref{fig:Load-data}, which includes data from the load cell
used to measure preloading through the jacking system.

Figure~\ref{fig:Load-data} shows the strain measurements converted
into load (based on strut dimensions and the elastic modulus) and
compared with the output of a load cell deployed on the strut and
the VW estimated load. The load cell clearly shows a diurnal pattern
that is in sync with the WBF strain measurements. These patterns are
most likely an effect of temperature variation. The VW measurements
show little or no variation with temperature, and have probably been
adjusted to compensate for temperature.

\section{Lessons from the deployments and open challenges\label{sec:Lessons-from-the}}

The ReStructure deployments and the subsequent data analysis revealed
a number of open questions for the WSN community as well as providing
several insights into more practical aspects of the implementation
and deployment that can increase the success of future projects if
catered for.

\subsection{Practical insights}
\begin{description}
\item [{Sensor~node~calibration}] The current sensor node requires manual
calibration on site, where environmental conditions can be extreme.
Setting up and calibrating nodes in situ also requires a wired connection
to the node. A wireless auto-calibration procedure (prompted remotely
by the deployer of the network) for nodes once they have been installed
could make the process faster and easier; alternatively, the balancing
process could be ignored and offsets applied during data processing.
\item [{Sensor~node~communication}] The current sensor network communication
technology uses 802.15.4 compatible radios transmitting at $2.4\GHz$.
It is possible that using a sub-1GHz frequency would increase the
range and stability of the multi-hop low power network. A platform
such as the Zolertia Remote~%
\footnote{http://zolertia.io/product/hardware/re-mote%
}  provides both $2.4\GHz$ and sub-$1\GHz$ radios, which could enable
side-by-side tests. Unfortunately, in the current deployment we have
been unable to perform a thorough analysis of the network operations
for dual-band radios. ReStructure performed well when deployed over
three strut levels, however the distance is only tens of metres. Intuitively,
short transmission distance is not an issue in a sufficiently dense
network (e.g., two nodes per strut). However, we did not deploy enough
sensors to see how the network scales. 
\item [{Sensor~node~packaging}] The current sensor node is rated to IP65.
We had however encountered instances of water ingress and one node
had water damage during severe rainstorms. Improving the IP rating
to IP67 or IP68 will ensure robustness.
\item [{Gateway}] The current gateway design makes use of batteries that
must be replaced every 30 days. A better design would be to add a
solar panel to extend battery life. In addition, the addition of a
sensor to measure and transmit battery charge level would have allowed
better scheduling of maintenance visits.
\end{description}

\subsection{Open design challenges}
\begin{enumerate}
\item More reliable, lower maintenance networks would be necessary to meet
performance requirements for accurate on line modelling and prediction
of load in TERS; methods for identifying the optimal sampling rate,
data transmission frequency and assessing connectivity changes can
help meet this aim (possibly thorough analysis of network activity
on various sites and evaluating the effects of evolving excavations).
Observations of network behaviour at scale is also necessary to inform
future developments. Open questions are to do with 1) the number of
gateways needed for best performance, 2) adding in redundancy, 3)
controlling the nodes and 4) implementing more dynamic operation of
the network at run-time so that data may be acquired at higher rates
when events of interest occur, such as strut pre-loading or excavation.
This is important since the loading is expected to be static most
of the time (excepting cyclic temperature variations that cause thermal
loading).
\item Lower energy networks are needed to meet the longevity requirements---the
ReStructure energy benchmarking section showed that 38\% of a nodes
energy requirement is for sensing, and 61\% is for Contiki Collect
operation (i.e. sending/forwarding packets and media access). To reduce
the energy requirement of a node we would need to i) user lower power
circuitry for the strain gauge signal conditioning hardware, and ii)
use a (potentially) lower power network protocol, such as TDMA (Time
division multiple access), would reduce the energy cost associated
with listening. 
\item Methods for detecting and compensating for thermal loading are needed
in order to make strain gauges a viable solution for cheap and dense
strain measurement.
\end{enumerate}

\section{Literature review\label{sec:Literature-review}}

A number of works in the literature have inspired and guided this
project, as well as providing baseline expectations for the deployments.
Specifically, we looked at related applications of WSNs (platforms,
packaging, and radio connectivity) for monitoring structural health.

Past research into monitoring of civil structures has attempted to:
integrate experimental and numerical modelling of seismic damage in
a three storey building (lab scale model)~\cite{Isidori}; track
performance of excavations during construction~\cite{McLandrich};
devise new sensing modalities applicable to structural health monitoring~\cite{Annamdas};
and assess the potential WSNs have for this field~\cite{Harms,Sun}.
While several recent works discuss aspects of WSN design for structural
monitoring~\cite{Galbreath,Jian16,Zhou16}, there have been few\emph{
long-term} (i.e., running over several years) research projects related
to the development and deployment of low-power wireless sensor network
systems on bridges and tunnels. Table~\ref{tab:notable-projects}
shows a breakdown of notable projects surveyed, including one recent,
but less developed work.

\begin{table}
\centering{}\protect\caption{Projects surveyed\label{tab:notable-projects}}
\begin{tabular}{>{\centering}p{0.3\columnwidth}>{\centering}p{0.3\columnwidth}c}
Institution/Department & Project & Duration\tabularnewline
\hline 
The University of Illinois at Urbana Champaign (UIUC) & Illinois Structural Health Monitoring Project (ISHMP) & 2002--\tabularnewline
The University of Michigan (U of M) & Yeondae Bridge, Grove Street Bridge, Guendam Bridge & 2004/5\tabularnewline
The University of California Berkeley (UCB) & Structural Health Monitoring of the Golden Gate Bridge & 2004--2006\tabularnewline
Cambridge University and Imperial College London & Smart Infrastructure: Monitoring and Assessment of Civil Engineering
Infrastructure & 2006--2009\tabularnewline
Swiss Federal Laboratories for Materials Testing and Research & Railway bridge monitoring & 2014--\tabularnewline
\hline 
\end{tabular}
\end{table}

\begin{itemize}
\item In 2004/5, Lynch \emph{et al}.~\cite{Lynch2006} performed two deployments
(2 days each) of a network of 14 custom sensor nodes (distance between
nodes of $5.75$--$19\m$, over $46\m$), measuring acceleration/vibration
for a concrete box girder bridge in Guemdang, Korea.
\item In 2006, a UCB team carried out a deployment of a 64-node multi-hop
network of MicaZ nodes ($2.4\GHz$) on the Golden Gate Bridge, sampling
vibration/acceleration at $1\kHz$ (July 2006 to October 2006, with
batteries changed in September 2006)~\cite{Kim2007}. Nodes were
spaced $46\m$ apart (close to the limit of communication). The authors
report average radio reception ranges around $30\m$ with a bi-directional
antenna, decreasing to around $15\m$ in the presence of obstructions.
The work informed our design choices particularly as the nodes placement
was similar and provided expectation for transmission ranges in the
excavation environment.
\item As part of the EPSRC WINES project, Cambridge University has had several
successful Structural Health Monitoring (SHM) deployments, including
a 12-node WSN installed on the Humber Bridge (approx. $60\m$ distance
from sink to furthest node), a 7-Node deployment on a concrete road
bridge (approx.~$30\m$) and a 26-node node deployment on the Jubilee
line on the London Underground (approx.~$180\m$ coverage). A key
finding is the need for connectivity testing prior to deployment and
further routing algorithms design that ensure fast convergence to
speed up deployment~\cite{Stajano2009a}.
\item As part of the Illinois Structural Health Monitoring Project~\cite{rice2010flexible},
the Jindo Bridge in South Korea has had extended deployment of sensors---70
in 2009, increased to 113 in 2011. Wind speed, acceleration and strain,
were measured across a heterogeneous network. Multiple aspects of
this deployment have been documented: Jang \emph{et al}.~\cite{jang2010structural}
describe the hardware and physical deployment including details of
the 4 networks (3 single hop networks of 23 nodes covering $175\m$,
and 1 multi-hop network) and emphasise that antenna placement is key
to increased transmission ranges, while Nagayama \emph{et al}.~\cite{nagayama2010reliable}
evaluated the use of multi-channel transmission to increase the available
throughput of the network in high data rate applications.
\item In 2014, Feltrin \emph{et al}.~\cite{Feltrin} developed and deployed
a custom sensor node for railway bridges to predict fatigue from train
traffic. Their custom sensor node was based on the TI MSP430 SoC,
and a sub-GHz radio. The deployment consisted of 7 nodes deployed
for 13 days on a railway bridge in November 2014. Over the 13 days
they achieved a PDR > 90\% through an event-based system design. Similar
to the ReStructure nodes, Feltrin \emph{et al}.~\cite{Feltrin2}
note the large power requirement for sensing. They solve this issue
by using low power components for their strain gauge signal conditioning
hardware.
\end{itemize}
A variety of platforms were used in these projects. Commonly, wireless
sensor nodes were constructed as a combination of i) an off-the-shelf
radio / microprocessor platform and ii) a custom I/O board to provide
the relevant functionality for the motivating application. We also
follow this approach in this work. Software modules and operating
systems are readily available for existing sensor node platforms (e.g.,
TinyOS and Contiki), however these platforms rarely provide the specialised
on-board signal processing chains that are required for vibration,
strain or other structural measurements. From deployments listed,
only Lynch~\cite{NewNarda} developed a custom hardware platform,
which is called the Narada. 

Radios transmitting on the $2.4\,\mathrm{GHz}$ ISM band (TelosB,
MicaZ and Imote) appear to be most common for structural monitoring
projects, with the exception of Hao \emph{et al.~}\cite{Hao} who
also used Mica2 ($433\MHz$), and Lynch \emph{et al.}~\cite{NewNarda}
where a custom node with a $900\MHz$ radio was used. The platform
of direct relevance here is the Imote2 based SHM platform~\cite{rice2010flexible}.
This is a stackable board configuration that modularises the sensor
node, data acquisition and signal conditioning to service a variety
of applications. The SHM-S board is used to measure strain, and has
undergone several stages of development to refine the circuitry and
operation. It features built-in resistors to perform shunt calibration
(i.e. fake strain values to test input/output range of the acquisition
system) and a digitally controlled potentiometer to enable auto-balancing
of a strain gauge in response to a system command. This type of functionality
is desirable for a well-tested board that has undergone significant
development but is over-complicated for a prototype strain circuit.

Radio connectivity and packaging impact the success of all practical
deployments and are thus discussed in more detail below.
\begin{itemize}
\item In controlled experimentation with the Imote / CC2420 radio Rice~\cite{Rice2010}
demonstrated that the use of an external, omnidirectional antenna
(the Antenova Titanis) had a significant effect in terms of Received
Signal Strength in an anechoic chamber when compared to an integrated/on
board antenna. Subsequent field tests examining line of sight transmission
ranges demonstrated that this antenna was capable of transmitting
around 3 times the distance of the on-board chip antenna ($300\m$
compared to $100\m$). The authors also found the effect of transmitting
through a steel / concrete floor (between floors in a University building
at a distance of $6\m$) produced only a 10\% reduction in reception
rate when using the external antenna, compared to line of sight transmission
between floors. Due to the improvement in reception rate using the
external antenna, the same antenna has been integrated to the ReStructure
node. However, we did not perform any comparison tests to assess performance. 
\item An investigation in 2010~\cite{kim2010rapid} used 20 Narada nodes
in an open parking lot to investigate the benefit of i) a custom amplifier
on the standard $2.4\GHz$ radio ($10\dB$ amplification) and ii)
directional antennas. With a standard antenna, maximum distance achievable
is between $200\m$ (no amplification) and $300\m$ (amplification).
With the directional antenna, the maximum distance increases to $500\m$
(no amplification) and $600\m$ (amplification). The power amplification
has a significant cost to the sensor node power budget and so represents
a design decision for specific applications, as does the directional
antenna, which limits the configuration of the sensor network. Since,
micro-benchmarking showed that the nodes would not meet the one year
lifetime requirement, and we could not access the nodes post-deployment,
the ReStructure node does not include a signal amplifier. Further
as the deployment site was ever evolving we settled for an omni-directional
antenna.
\item In 2008, Hao \emph{et al}.~\cite{Hao} performed a study of radio
connectivity during the construction of Pasir Panjang MRT station
in Singapore, as part of a project related to an Integrated Dual Radio
Framework for sensor networks. This unpublished work examined the
difference in radio connectivity between MicaZ ($2.4\GHz$) and Mica2
($900\MHz$) sensor nodes. Unfortunately, the authors only deployed
sensors at one side of the excavation, and then only at the top. The
nodes performed a simple packet counting exercise, with individual
nodes transmitting a burst of packets every 11 minutes while its neighbours
counted the number of packets received. The longest period tested
was for 5.5 days. The authors found positioning of the antenna is
vital for connectivity. When the nodes were simply placed on the struts,
the $2.4\GHz$ radio had slightly better performance (this result
was duplicated in the lab with the nodes placed on the floor). However,
when the antenna was raised, the opposite was true. In Feltrin's~\cite{Feltrin}
project they used a sub-GHz radio in their custom sensor to achieve
better signal propagation as no nodes had line of sight. The ReStructure
nodes used a $2.4\GHz$, however during range tests we discovered
that communication range was a limiting factor. In future work, we
aim to compare $2.4\GHz$ and sub-GHz radios. The choice of the radio
frequency will affect the resulting network infrastructure. For example,
longer transmission ranges may reduce the number of gateways required.
\item A variety of different approaches have been used to attach sensors
onto the structure being monitored. In one bridge monitoring project,
enclosures and antennas were attached with G-clamps~\cite{Kim2007}.
In the Jindo bridge project~\cite{rice2010flexible}, magnets were
used to connect the sensor nodes to the underside of the bridge along
with curved brackets where nodes needed to be attached to support
poles. In both projects, plastic enclosures were used.
\end{itemize}

\section{Conclusions\label{sec:Conclusions}}

The paper demonstrates that WSN based systems can be effectively used
for monitoring load in TERS during excavation and construction processes.
Through the design, implementation and deployments reported, we showed
that, when well specified at design stage, WSNs:
\begin{itemize}
\item can operate, as required, in the field, unattended, and deliver data
to a remote user application;
\item can achieve consistently high data yield (observed between 81.3\%
to 91.1\% in normal operating conditions) on single and multi-hop
configurations, using $2.4\GHz$ radio, on active and dynamic construction
sites, under extreme climatic conditions and in face of obstructions
and interference common to excavation sites. Human error caused the
greatest proportion of data loss;
\item can evolve their network structure as the number of nodes increases
while construction proceeds and depth of excavation increases;
\item can be designed to function, on $2\times7800\,\mathrm{mAh}$ C cells,
for a duration of 6 months in sense-and-send mode, with 5 minute sampling
interval.
\end{itemize}
Further research is however needed in order for such systems to achieve
their full, large scale industrial potential:
\begin{itemize}
\item Solutions must be found to better compensate for apparent strain in
the measurement system without removing the effects of thermal loading;
ideally this would be done through in-field calibration under no-load
conditions (before pre-loading), or using manufacturer provided polynomial
coefficients based on batch testing during production; 
\item to enable future systems to minimise their power consumption and extend
battery life, they will need network protocol and hardware optimisation
as well as strategies for adaptive sampling;
\item future systems must be easily deployable and self-calibrating to minimise
deployment time in extreme deployment environments;
\item understanding the impact on wireless communication and data collection
capability of a full-scale deployment on a TERS, grown throughout
the entirety of its life-cycle (a year or more);
\item real-time integration of strain data with structural soil models also
remains an open are of research that is worthy of consideration by
the community--deriving tight requirements for the WSN from the model
side and piloting full end-to-end systems is key to ensure full benefit
is obtained from the WSN technologies
\end{itemize}

Increasing the density, timeliness and frequency of relatively cheap
strain measurements on TERS can give new insight into the design and
construction process. We can foresee smart struts where instrumentation
is embedded into the support. At this stage the challenges lie in
mass collection and integration of these measurements to motivate
such a move to intelligent construction infrastructure.

\bibliographystyle{plain}
\bibliography{references}

\end{document}